\documentclass[12pt]{article}
\usepackage{amssymb}

\input{tcilatex}

\begin{document}

\bigskip \baselineskip0.8cm \textwidth16. cm

\begin{center}
\textbf{\ Reentrant Phase Transitions of the Blume-Emery-Griffiths Model for
a Simple Cubic Lattice on the Cellular Automaton}

N. Sefero\u{g}lu$^{\ast }$ and B. Kutlu$^{\ast \ast }$

$^{\ast }$ Gazi \"{U}niversitesi, Fen Bilimleri Enstit\"{u}s\"{u}, Fizik
Anabilim Dal\i ,

Ankara, Turkey

$^{\ast \ast }$ Gazi \"{U}niversitesi, Fen -Edebiyat Fak\"{u}ltesi, Fizik B%
\"{o}l\"{u}m\"{u},

06500 Teknikokullar, Ankara, Turkey

e-mail: nurguls@gazi.edu.tr

e-mail: bkutlu@gazi.edu.tr

\textbf{\ }

\bigskip
\end{center}

\textbf{Abstract}

The spin-1 Ising (BEG) model with the nearest-neighbour bilinear and
biquadratic interactions and single-ion anisotropy is simulated on a
cellular automaton which improved from the Creutz cellular automaton(CCA)
for a simple cubic lattice. The simulations have been made for several sets
of parameters $K/J$ and $D/J$ in the $-3<D/J\leq 0$ and $-1\leq K/J\leq 0$
parameter regions. The re-entrant and double re-entrant phase transitions of
the BEG model are determined from the temperature variations of the
thermodynamic quantities ($M$, $Q$ and $\chi $ ). The phase diagrams
characterizing phase transitions are compared with those obtained from other
methods.


*Corresponding author

\textbf{1 . Introduction}

The Blume-Emery-Griffiths (BEG) model$^{1}$ was originally introduced in
order to explain the phase separation and superfluidity in the He$^{\text{3}%
} $-He$^{\text{4}}$ mixtures. Subsequently, the model was used in the
description of a variety of different physical phenomena such as
multicomponent fluids$^{2}$, microemulsions$^{3}$, and semiconductors alloys$%
^{4}$, etc.

The Hamiltonian of the BEG model is given by,

\begin{equation}
H_{I}=J\sum_{<ij>}S_{i}S_{j}+K\sum_{<ij>}S_{i}^{2}S_{j}^{2}+D%
\sum_{i}S_{i}^{2}
\end{equation}%
where $s_{i}=\pm 1,0$ and $<ij>$ denotes summation over all
nearest-neighboring (nn) spin pairs on a simple cubic lattice. The
parameters of $J$ and $K$ are the bilinear and biquadratic interaction
energies, respectively and $D$ is the single-ion anisotropy constant. The
three-dimensional BEG\ model has been extensively studied by different
techniques, using the mean-field approximation (MFA)$^{(1,5-7)}$,
effective-field theory$^{(8-11)}$, two-particle cluster approximation (TPCA)$%
^{12}$, Bethe approximation$^{13}$, high-temperature series expansion$^{14}$%
, renormalization group theory$^{15}$, Monte Carlo simulations$^{(13,16-17)}$%
, linear chain approximation$^{(18,19)\text{ }}$and cellular automaton$%
^{(20,21)}$

In this paper we studied the three-dimensional BEG\ model using an improved
heating algorithm from the Creutz Cellular Automaton (CCA) for simple cubic
lattice. The CCA algorithm is a microcanonical algorithm interpolating
between the canonical Monte Carlo and molecular dynamics techniques on a
cellular automaton, and it was first introduced by Creutz$^{22}$. In the
previous papers$^{(20,21,23-26)}$, the CCA algorithm and improved algorithms
from CCA were used to study the critical behavior of the different Ising
model Hamiltonians on the two and three-dimensions. It was shown that they
have successfully produced the critical behavior of the models.

The BEG model has a complicated phase diagrams and has several kinds of
phase transitions, such as re-entrant and double re-entrant transitions for
some values of the model parameters on the three-dimensional lattices.
However, there exists the differences between the phase diagrams in the ($%
d=D/J$, $k=K/J$, $t=k_{B}T/zJ$) plane is obtained for certain parameter
values$^{(6,12,13,16)}$, where $z$ is the coordination number. Recently, TPCA%
$^{12}$ calculations of the three dimensional BEG model with the
ferromagnetic bilinear interaction show that the phase diagrams are
qualitatively different from the phase diagrams obtained with the MFA$^{6}$
and Bethe approximation$^{13}$ for the $k=-0.5$ parameter. In addition, the
order parameters exhibit a different critical behavior for the double
re-entrant phase transition. Thus, the alternative results are needed to
check the differences in results. The aim of the present paper is to study
re-entrant behaviors of the BEG model in the case of $J<0$ and to obtain the
phase diagrams for several sets of parameters $k$ and $d$ in the $-3<d\leq 0$
and $-1\leq k\leq 0$ parameter regions. For this purpose, the thermodynamic
quantities are computed \ on a simple cubic lattice with linear dimension $%
L=18$ and the finite critical temperatures are estimated from the maxima of
the susceptibility. The model is explained in Sec.2, the results are
discussed in Sec.3 and a conclusion is given in Sec.4.

\textbf{2. Model}

Three variables are associated with each site of the lattice. The value of
each sites are determined from its value and those of its nearest-neighbors
at the previous time step. The updating rule, which defines a deterministic
cellular automaton, is as follows: Of the three variables on each site, the
first one is Ising spin $B_{i}$. Its value may be $0$ or $1$ or $2$. The
Ising spin energy for the model is given by $Eq.1$. In $Eq.1$, $Si=Bi-1$.
The second variable is for momentum variable conjugate to the spin (the
demon). The kinetic energy associated with the demon, $H_{k}$, is an
integer, which equal to the changing in the Ising spin energy for the any
spin flip and its values lie in the interval (0, m). The upper limit of the
interval, m, is equal to 24J. The total energy

\begin{equation}
H=H_{I}+H_{K}
\end{equation}

is conserved.

The third variable provides a checkerboard style updating, and so it allows
the simulation of the Ising model on a cellular automaton. The black sites
of the checkerboard are updated and then their color is changed into white;
white sites are changed into black without being updated. The updating rules
for the spin and the momentum variables are as follows: For a site to be
updated its spin is changed one of the other two states with $1/2$
probability and the change is transferable to or from the momentum variable
associated with this site, such that the total energy $H$ is conserved, then
this change is done and the momentum is appropriately changed. Otherwise the
spin and the momentum are not changed.

For a given total energy the system temperature is obtained from the average
value of kinetic energy, which is given by:

\begin{equation}
<E>=\frac{\sum\limits_{n=0}^{m}ne^{-nJ/kT}}{\sum\limits_{n=0}^{m}e^{-nJ/kT}}
\end{equation}%
where $E=H_{K}.$ The expectation value in $Eq.3$ is a average over the
lattice and the number of the time steps. Because of the third variable, the
algorithm requires two time steps to give every spin of the lattice a change
to change. Thus, in comparison to ordinary Monte Carlo simulations, two
steps correspond to one full sweep over the system variables.

The heating algorithm is divided into two basic parts, initialization
procedure and the taking of measurements. In the initialization procedure,
firstly, all spins in the lattice sites take the ferromagnetic ordered
structure ($\left\uparrow {}\right\uparrow )$ and the kinetic energy per
site which is equal to the maximum change in the Ising spin energy for the
any spin flip is given to the certain percent of the lattice via the second
variables. This configuration is run during the 10.000 cellular automaton
time steps. At the end of the this step, the configuration in the ordered
structure at the low temperature is obtained. In the next steps last
configuration in the ordered structure has been chosen as a starting
configuration for the heating run. Rather than resetting the starting
configuration at each energy, it is used the final configuration at a given
energy as the starting point for the next. During the heating cycle, energy
is added to the system through the second variables ($H_{k}$) after
1.000.000 cellular automaton steps.

\textbf{3. Results and discussion}

The three-dimensional BEG model is simulated with the heating algorithm
which improved from the Creutz Cellular Automaton. The simulations were
executed on simple cubic lattice $L$x$L$x$L$ of the linear dimensions $L=18$
with periodic boundary conditions. The computed values of the quantities are
averages over the lattice and over the number of time steps (1.000.000) with
discard of the first 100.000 time steps during which the cellular automaton
develops. Hereafter we shall use the terminology of Ref. 13 for the
definition of phases: F, the ferromagnetic phase ($M\neq 0,$ $Q\neq 2/3$);
Q, the quadrupolar phase ($M=0,$ $Q\neq 2/3$).

In the ground state, the phase diagram of the BEG model on the ($d$, $k$)
plane for $-1\leq k\leq 0$ parameter region is shown in Fig.1. The region of
the perfect zero ordering and ferromagnetic ordering are separated by the
line$^{13}$ $k=-1-d/3$ represented with the solid line. The types of phase
transition (PT) obtained from the CCA calculations are also shown in Fig.1.
In the ferromagnetic ordering region, the $Q\rightarrow F$ \ phase
transitions occur according to the values of $k$ and $d$. However, the
re-entrant $Q\rightarrow F\rightarrow Q$ and double re-entrant $Q\rightarrow
F\rightarrow Q\rightarrow F$ phase transitions take place for some values of 
$k$ and $d$ near the phase boundary.

For the reproduce of the phase diagram using the CCA simulations, the
temperature variation of the order parameters ($M$ and $Q$) and the
susceptibility ($\chi $) are calculated at $d=0,-0.25,-1$ and $-2.5$ for
several $k$ values on the finite lattice with $L=18$. For $d=0$, the order
parameters and the susceptibility data are shown in Fig.2 (a) and (b). There
is the second order phase transition from quadrupolar phase to ferromagnetic
phase ($Q\rightarrow F$) at $d=0$ for $k\geq -1$. In this region, the
susceptibility data shows characteristic peaks at the critical temperature.
However for $k=-1$ the value of the magnetization at the lowest temperature
is much lower than the magnetization for the $k>-1$ values.

The order parameters $M$ and $Q$ obtained at $d=-0.25$ for different $k$
values are shown in Fig.3 (a) and (b). For $k\geq -0.90$, there is only one
second order $Q\rightarrow F$ phase transition in the system. In the
interval $-0.95\leq k<-0.90$, the re-entrant $Q\rightarrow F\rightarrow Q$
phase transitions take place and both of them are second order. At the same
time, the data of the susceptibility shows the two peaks which belong the $%
Q\rightarrow F$ and $F\rightarrow Q$ phase transitions in the $-0.95\leq
k<-0.90$ parameter region [Fig. 3(c)].

For $d=-1$, the temperature dependence of the order parameters and the
susceptibility are given in Fig.4. As seen from Fig.4, while the
second-order $Q\rightarrow F$ phase transition occurs for $k>-0.66$, the
double re-entrant $Q\rightarrow F\rightarrow Q\rightarrow F$ phase
transitions occur for $k=-0.66$, $-0.67$ and $-0.68$. For $k=-0.66$, only
the higher transition $Q\rightarrow F$ is second order and the others are
first order. At the same time, the susceptibility for the lower and middle
transitions show a sharp peaks at two different transition temperature
[Fig.4 (c)]. Nevertheless, for $k=-0.67$ and $-0.68$ only the lower phase
transition is first order because the order parameters exhibit a
discontinuous behavior and the susceptibility exhibit a jump. According to
these results, the middle phase transition $F\rightarrow Q$ in the double
re-entrant $Q\rightarrow F\rightarrow Q\rightarrow F$ phase transition can
be the first order and second order. This is in agreement with the results
obtained by the MFA$^{6}$ and \ TPCA$^{12}$ but disagreement with Bethe
approximation(BA)$^{13}$ which gives only the lower phase transition can be
first order.

In Fig.5, the order parameters and the susceptibility are shown for
different values of the $k$ parameter at $d=-2.5$. The re-entrant or double
re-entrant PT's are not seen for $d=-2.5$. While the $Q\rightarrow F$ phase
transition is second order for $k\geq -0.12$, the $Q\rightarrow F$ phase
transition is first order for $k<-0.12$. Nevertheless, the susceptibility
shows a characteristic behavior for $k\geq -0.12$ while a jump in the
susceptibility appears for $k<-0.12$.

Consequently, while the $Q\rightarrow F$ PT take place at $\ d=0$ for $k\geq
-1$, the re-entrant $Q\rightarrow F\rightarrow Q$ and double re-entrant $%
Q\rightarrow F\rightarrow Q\rightarrow F$ PT's occur at $d=-0.25,-0.5,-1$
and $-1.5$ for certain values of $k$. The representative phase diagrams
relevant to the these phenomena are shown in Fig.6. The phase diagrams
obtained at $d=0$ and $-1.5$ represent similar behavior with the Bethe
approximation results$^{13}$ illustrated in Fig.6.

Finally, we obtain the ($d$, $t$) phase diagram for $k=-0.5$ [Fig.7]. \ It
is quite different from the phase diagram obtained with different methods $%
^{(6,15)}$ at $k=-0.5$. In the phase diagram obtained with the MFA$^{6}$ and
renormalization group theory$^{15}$ for $k=-0.5$, the second order PT line
terminates at the critical end point E at the first order PT line and the
first order PT line terminates at the critical point C inside the
ferromagnetic phase. On the other hand, there is a tricritical point in the
phase diagram obtained with TPCA$^{12}$ for $k=-0.5$. As seen Fig.7, there
is no critical point C in the ferromagnetic phase and only a tricritical
point exists in the phase diagram with CCA calculations as in TPCA.

\textbf{4. Conclusion}

The BEG model is simulated using the heating algorithm of the cellular
automaton for simple cubic lattice. The CCA calculations show that the BEG\
model exhibits the re-entrant and double re-entrant phase transitions for
some values of\ the $k$ and $d$ parameters near the phase boundary as
expected. According to the Bethe approximation$^{13}$, the lower PT in the
double re-entrant $Q\rightarrow F\rightarrow Q\rightarrow F$ PT is of the
first order and the other PT's are of the second order. Contrast to this,
the MFA$^{6}$ and TPCA$^{12}$ calculations suggest that the middle PT can be
also the first order. The CCA results are compatible with the MFA and TPCA
results at this point. We have also reconstructed the phase diagrams in the (%
$d$, $k$, $t$) parameters space using the CCA simulations. The phase
diagrams in the ($k$, $t$) plane for $d=0$ and $-1.5$ are similar to the
phase diagrams obtained by the BA$^{13}$ and TPCA$^{12}$ at those
parameters. Although the MFA$^{6}$ and renormalization group theory$^{15}$
predict that the ($d$, $t$) phase diagram for $k=-0.5$ has an end point at
the first order phase transition line and a critical point in the
ferromagnetic phase, the CCA calculations for $k=-0.5$ indicate that there
is only a tricritical point as the results of \ TPCA$^{12}$.

\textbf{Acknowledgements}

This work is supported by a grant from Gazi University(BAP:05/2003-07).

\textbf{References}

[1] M. Blume, V.J. Emery and R.B. Griffiths, Phys. Rev. A 4 (1971) 1071.

[2] J. Lajzerowicz, J. Sivardiere, Phys. Rev. A 11 (1975) 2079; J.
Sivardiere, J. Lajzerowicz, Phys. Rev. A 11 (1975) 2090.

[3] M. Schick, W. H. Shih, Phys. Rev. B 34 (1986) 1797.

[4] K. E. Newman, J. D. Dow, Phys. Rev. B 27 (1983) 7495.

[5] H.H. Chen and P. M. Levy, Phys. Rev.B 7, (1973) 4267.

[6] W. Hoston and A. N. Berker, Phys. Rev. Lett. 67 (1991) 1027.

[7] J. A. Plascak, J. G. Moreira and F. C. Sa Barreto, Phys. Lett. A 173
(1993) 360.

[8] T. Kaneyoski and E. F. Sarmento, Physica A 152, (1988) 343.

[9] J.W. Tucker, J.\ Magn. Magn. Mater 87 (1990) 16.

[10] K. G. Chakraborty, Phys. Rev. B 29 (1984) 1454.

[11] A. F. Siqueira and I. P. Fittipaldi, Phys. Rev. B 31 (1985) 6092.

[12] O. R. Baran and R. R. Levitski, Phys. Rev. B 65 (2002) 172407.

[13] K. Kasono and I. Ono, Z. Phys. B: Cond. Matter 88 (1992) 205.

[14] D.M. Saul, M. Wortis and D. Stauffer, Phys.Rev. B 9 (1974) 4964.

[15] R. R. Netz and A. N. Berker, Phys.Rev. B 47 (1993) 15019.

[16] O. F. Alcantara Bonfirm and C. H. Obcemea, Z. Phys. B: Cond. Matter 64
(1986) 469.

[17] D. Wimgert and D.Stauffer, Physica A 219 (1995) 135.

[18] J. A. Plascak, N. P. Silva, Phys. Stat. Sol. (b) 110 (1982) 669.

[19] E. Albayrak, M. Keskin, J. Magn. and Magn. Materials 203 (2000) 201.

[20] B. Kutlu, A. \"{O}zkan, N. Sefero\u{g}lu, A. Solak and B. Binal, Int.
J. Mod. Phys. C 16 (2005) 1933.

[21] A. \"{O}zkan, N. Sefero\u{g}lu and B. Kutlu, Physica A 362 (2006) 327.

[22] M. Creutz, Phys. Rev. Lett. 50 (1983) 1411.

[23] B. Kutlu, Physica A 234 (1997) 807.

[24] B. Kutlu, Int. J. Mod. Phys. C 12, (2001) 1401

[25] B. Kutlu, Int. J. Mod. Phys. C 14, (2003) 1305

[26] N. Aktekin, Annual Reviews of Computational Physics VII, ed. D.
Stauffer (World Scientific, Singapore, 2000), pp. 1-23.

\textbf{Figure Captions}

Fig.1. The phase diagram ($d$, $k$ plane) in the ground state and the types
of phase transitions obtained on a simple cubic lattice.

Fig.2. The temperature dependence of (a) the order parameters $M$ and $Q$,
(b) the susceptibility $\chi $ for $d=0$.

Fig.3. The temperature dependence of (a) the order parameter $M$, b) $Q$,
(c) the susceptibility $\chi $ for $d=-0.25$.

Fig.4. The temperature dependence of (a) the order parameters $M$, b) $Q$,
(c) the susceptibility $\chi $ for $d=-1$.

Fig.5. The temperature dependence of (a) the order parameters $M$ and $Q$,
(b) the susceptibility $\chi $ for $d=-2.5$.

Fig.6. The phase diagrams in the ($k$, $t$) plane for values of the $d=0$,
-0.25, -0.5, -1 and -1.5. The closed and open symbols represent the
second-order and first-order phase transitions for the CCA calculations,
respectively. The dashed and thick dashed lines indicate the second-order
and first-order phase transitions for the Bethe approximation (BA)$^{13}$ \
at the $d=0$ and $-1.5$ parameter values.

Fig.7. The phase diagram in the ($d$, $t$) plane for $k=-0.5$. The solid
line and dotted line represent the second-order and first-order phase
transitions for TPCA$^{12}$ calculations. The closed and open squares
represent the second-order and first-order phase transitions for CCA
calculations, respectively.\bigskip 

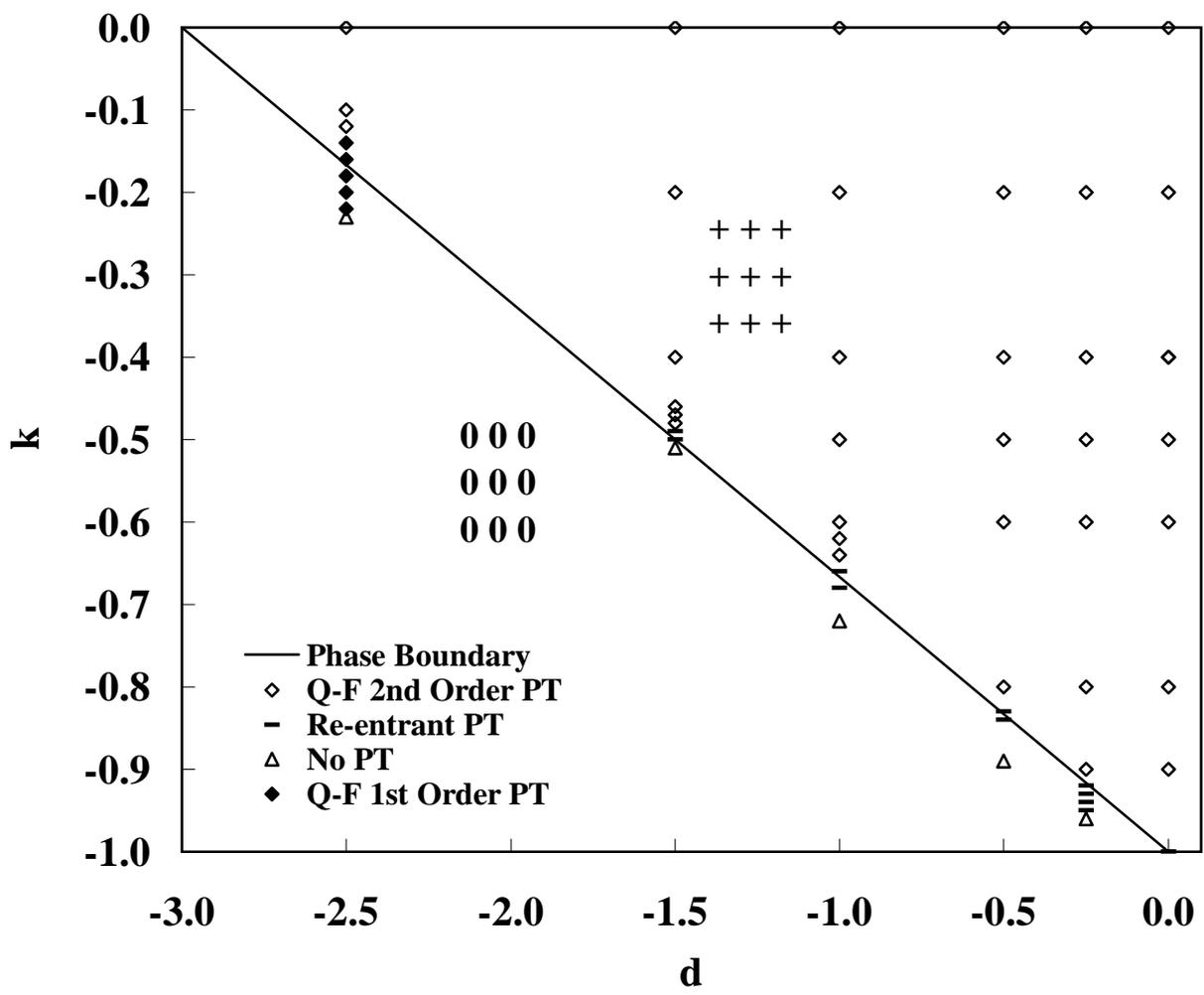
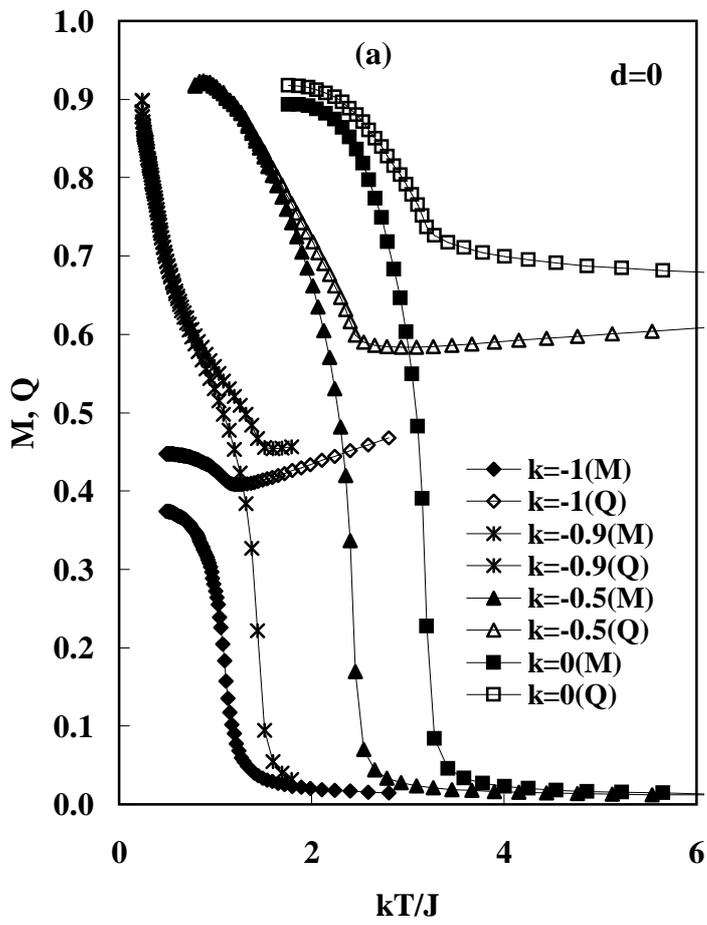
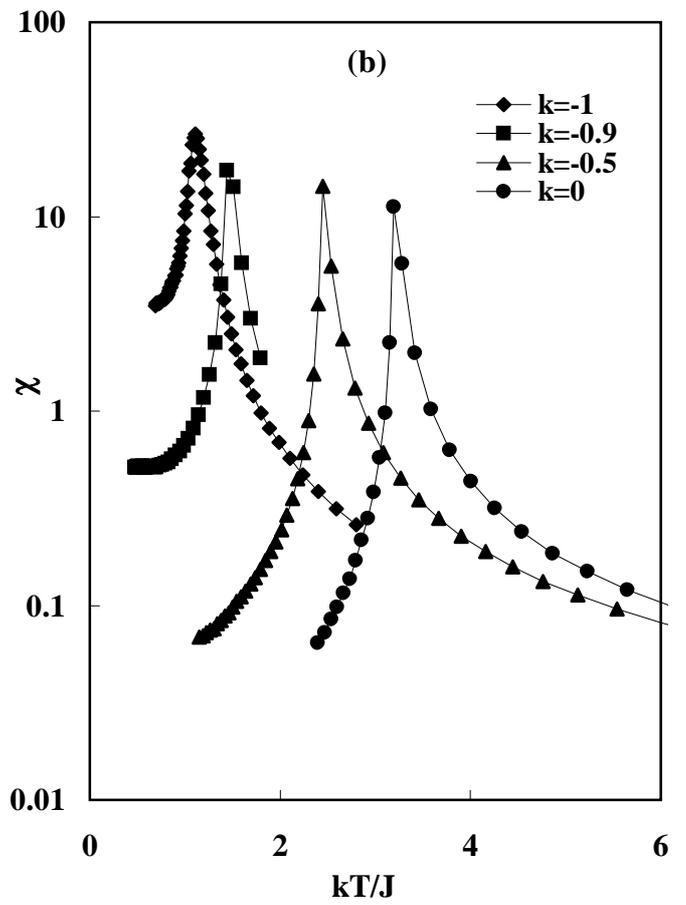
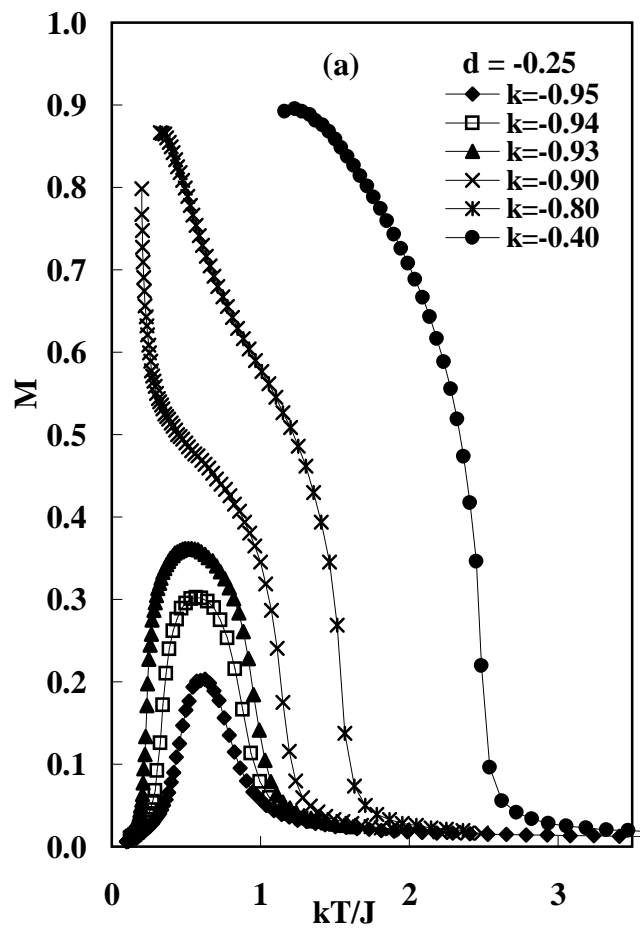
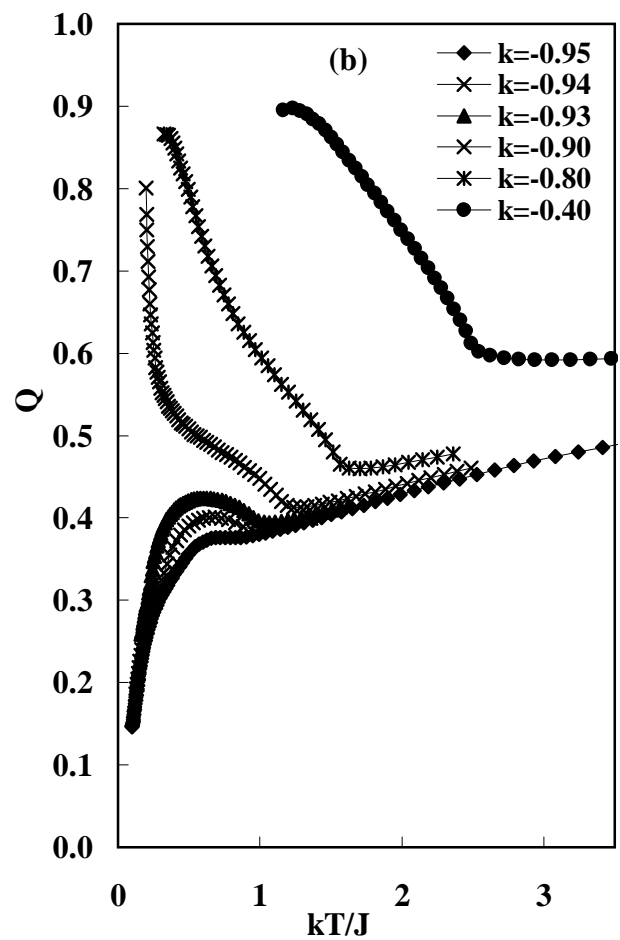
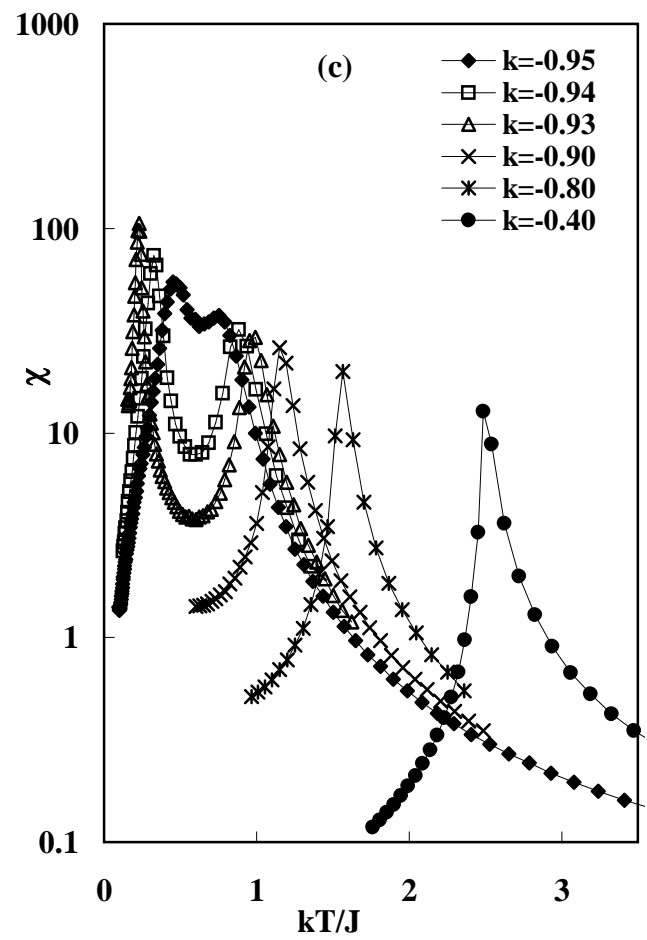
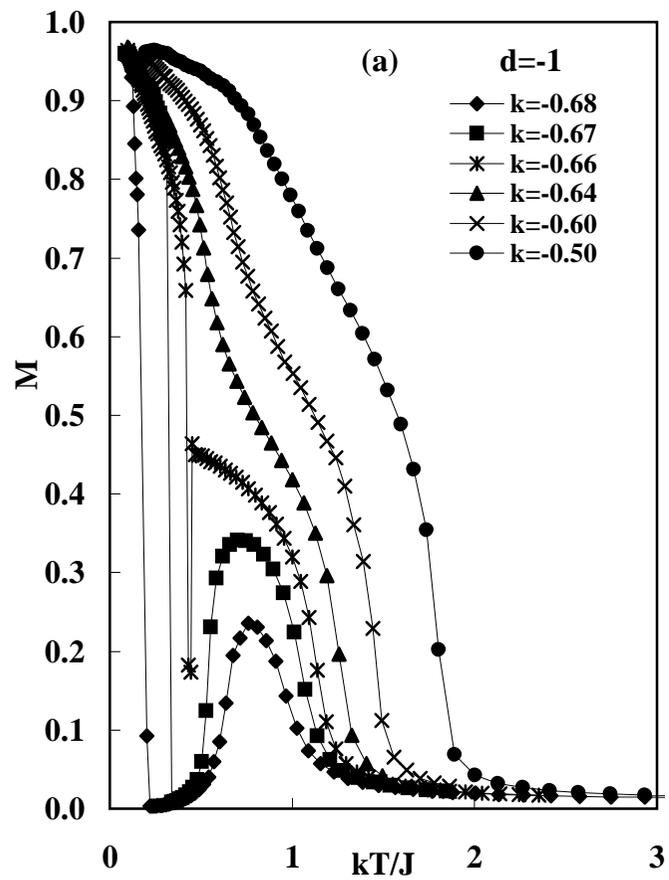
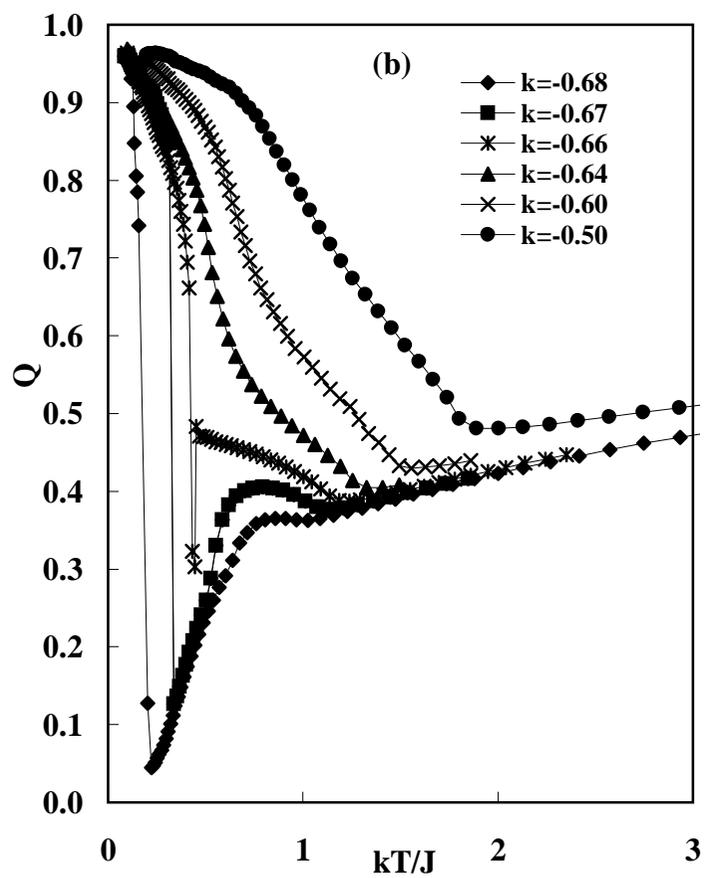
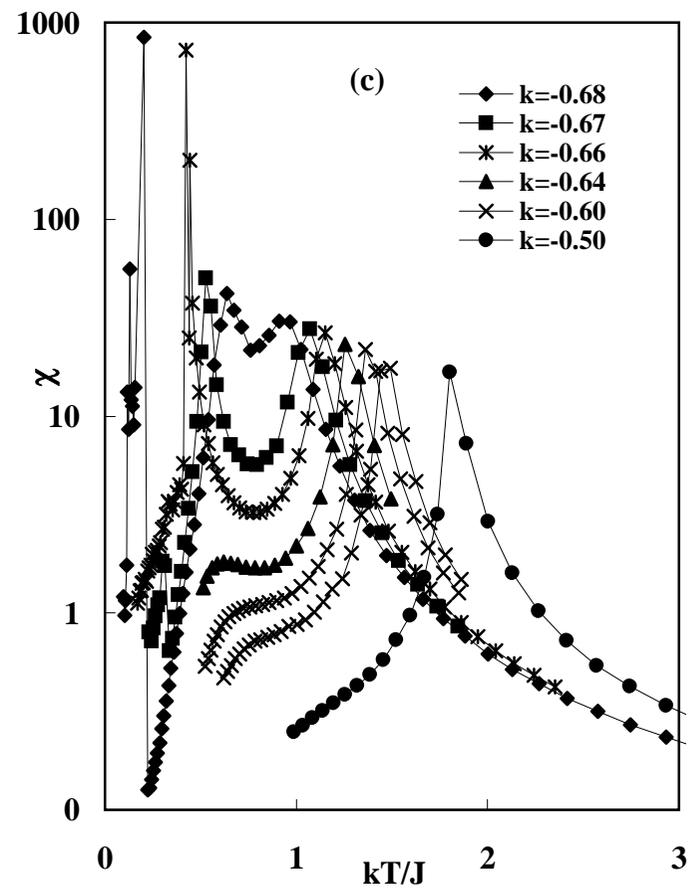
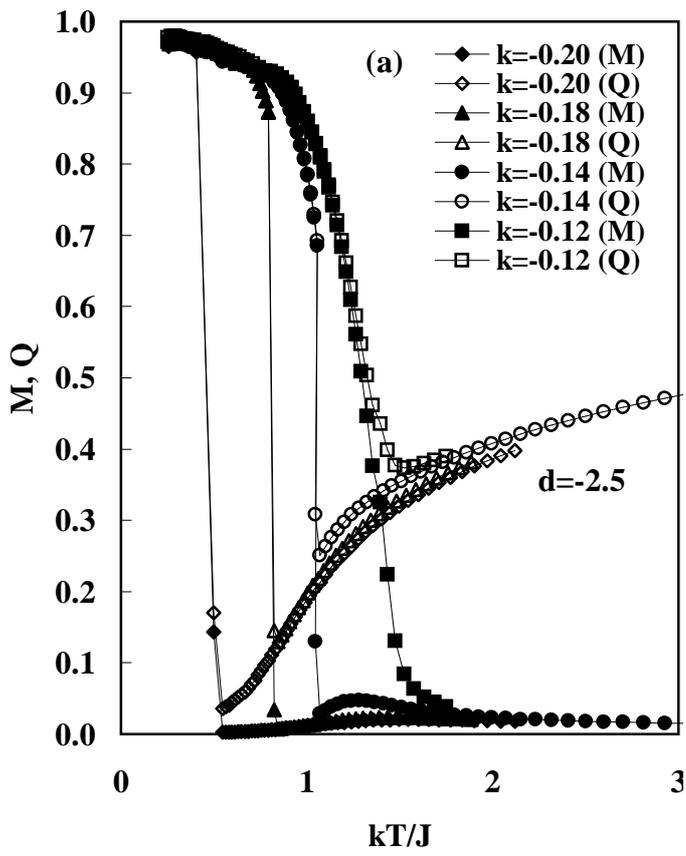
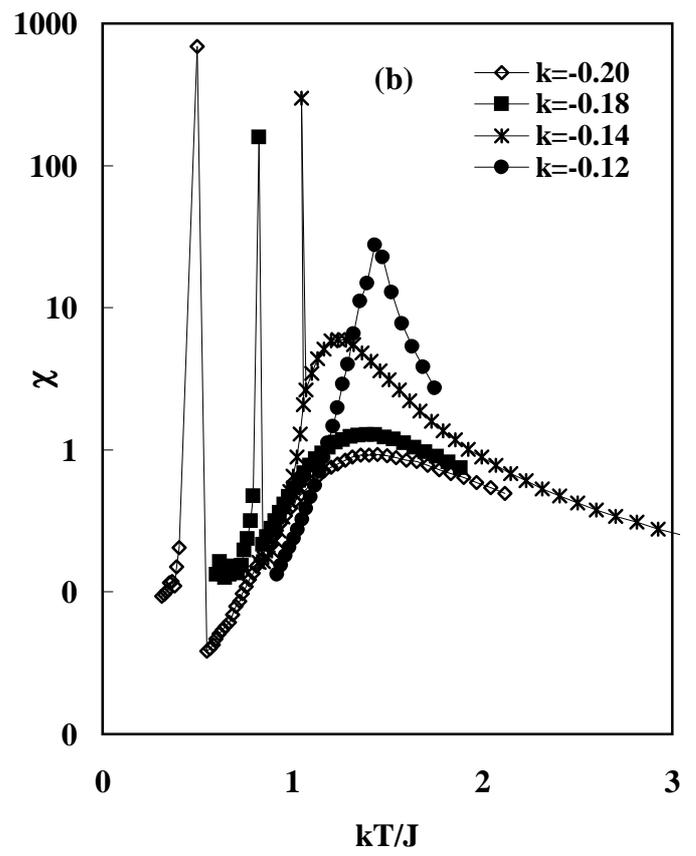
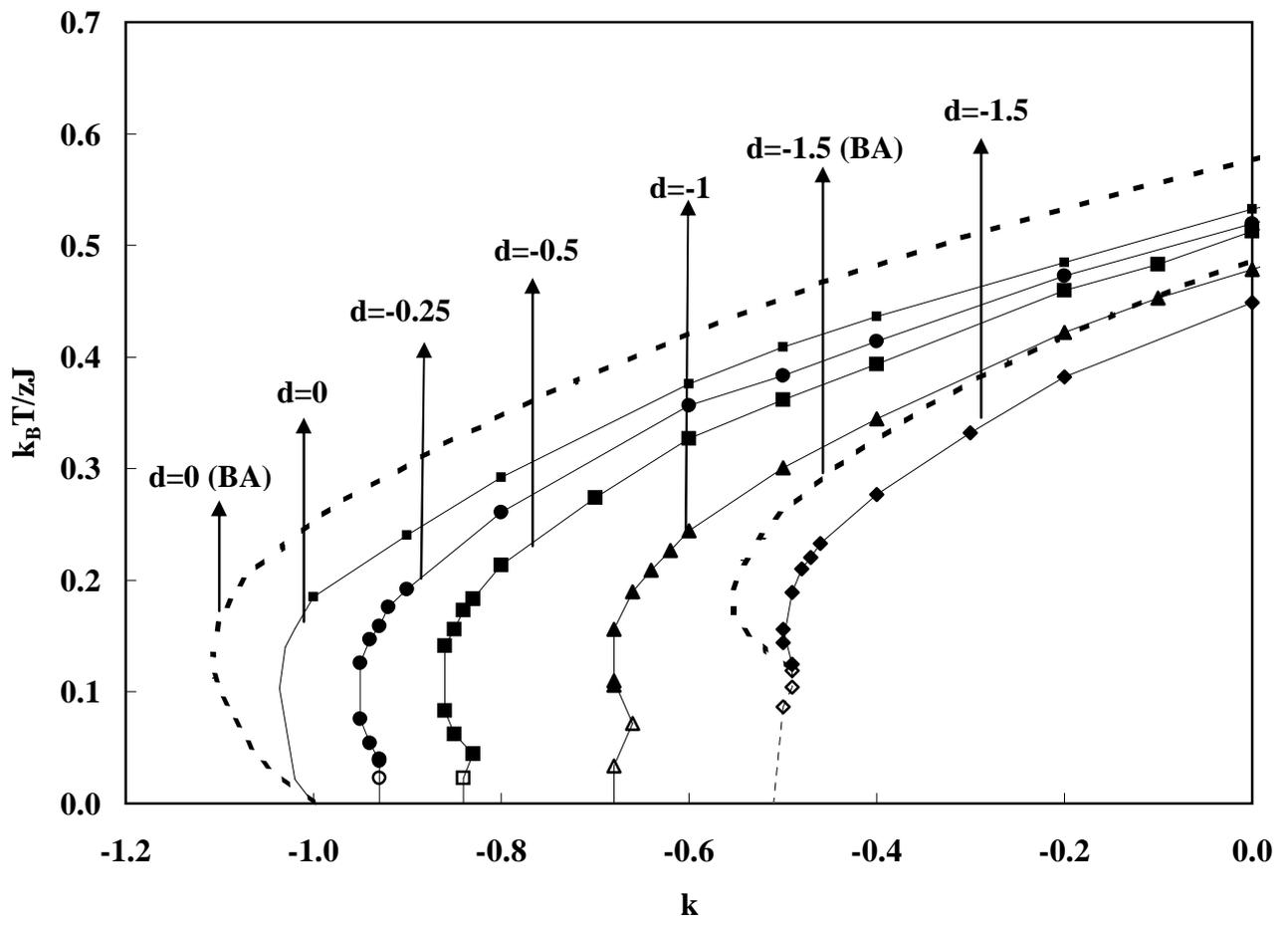
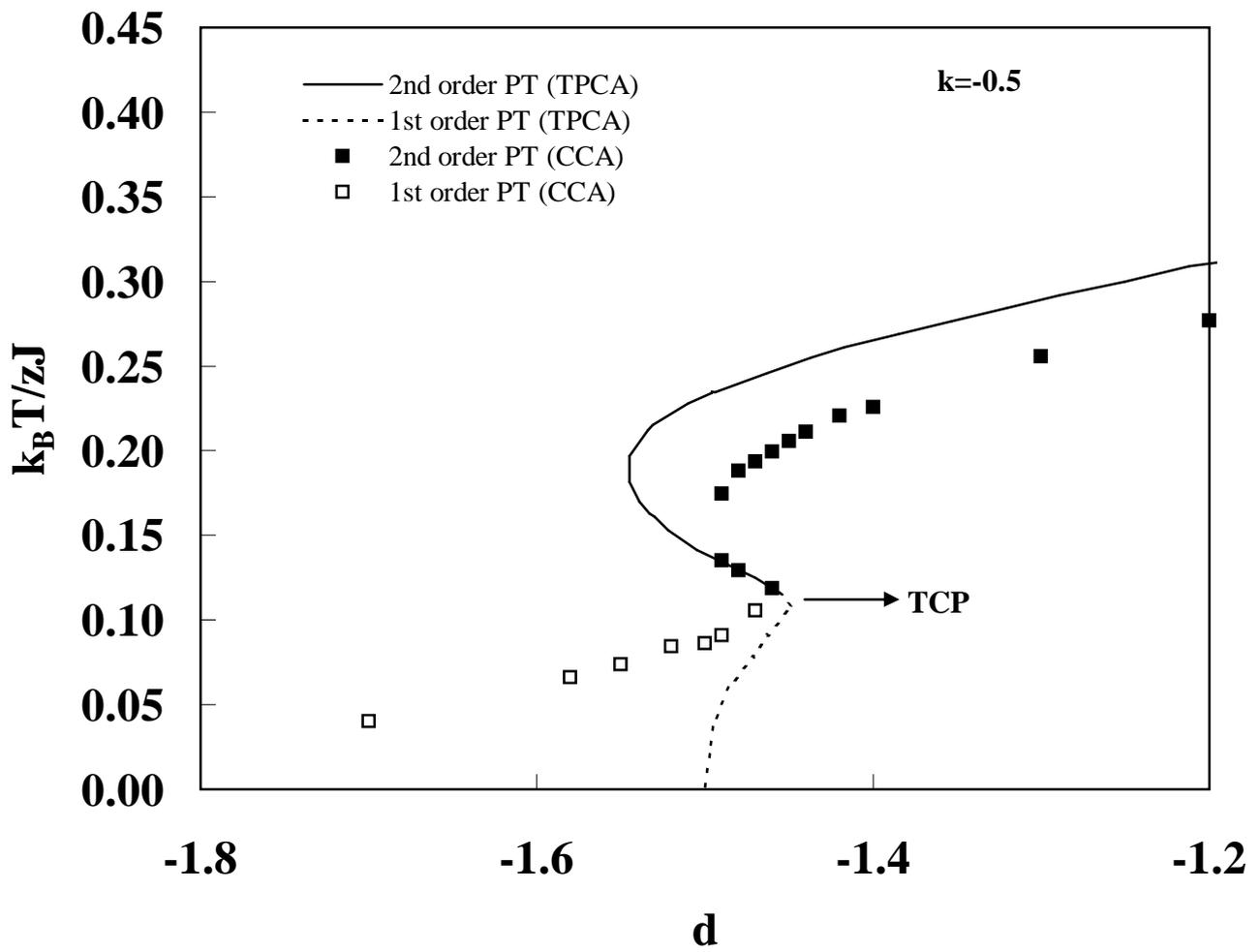
\end{document}